\documentstyle{AGILEpsr}

\input epsf.sty
\input epsfig.sty
\input psfig.sty

\def \AAP #1 #2 {{\em Astron. Astrophys.\/} {\bf #1}, #2}
\def \AAL #1 #2 {{\em Astron. Astrophys. Lett.\/} {\bf #1}, L#2}
\def \AAR #1 #2 {{\em Astron. Astrophys. Rev.\/} {\bf #1}, #2}
\def \AAS #1 #2 {{\em Astron. Astrophys. Suppl. Ser.\/} {\bf #1}, #2}
\def \AJ #1 #2 {{\em Astron. J.\/} {\bf #1}, #2}
\def \ANNREV #1 #2 {{\em Ann. Rev. Astron. Astrophys.\/} {\bf #1}, #2}
\def \APJ #1 #2 {{\em Astrophys. J.\/} {\bf #1}, #2}
\def \APL #1 #2 {{\em Astrophys. Lett.\/} {\bf #1}, L#2}
\def \APJL #1 #2 {{\em Astrophys. J. Lett.\/} {\bf #1}, L#2}
\def \APJS #1 #2 {{\em Astrophys. J. Suppl.\/} {\bf #1}, #2}
\def \APSS #1 #2 {{\em Astrophys. Space Sci.\/} {\bf #1}, #2}
\def \ASR #1 #2 {{\em Adv. Space Res.\/} {\bf #1}, #2}
\def \BAIC #1 #2 {{\em Bull. Astron. Inst. Czechosl.\/} {\bf #1}, #2}
\def \JSQRT #1 #2 {{\em J. Quant. Spectrosc. Radiat. Transfer\/} {\bf #1}, #2}
\def \MN #1 #2 {{\em Mon. Not. R. Astr. Soc.\/} {\bf #1}, #2}
\def \MEM #1 #2 {{\em Mem. R. Astr. Soc.\/} {\bf #1}, #2}
\def \PLR #1 #2 {{\em Phys. Lett. Rev.\/} {\bf #1}, #2}
\def \PASJ #1 #2 {{\em Publ. Astron. Soc. Japan\/} {\bf #1}, #2}
\def \PASP #1 #2 {{\em Publ. Astr. Soc. Pacific\/} {\bf #1}, #2}
\def \NAT #1 #2 {{\em Nature\/} {\bf #1}, #2}
\def \SAIT #1 #2 {{\em Mem.\ Soc.\ Astron.\ It.\/} {\bf #1}, #2}
\def \MESS #1 #2 {{\em The Messenger\/} {\bf #1}, #2}
\def \ASTRNACH #1 #2 {{\em Astron. Nach.\/} {\bf #1}, #2}
\def \AGPSR #1 #2 {{\em ASI Special Publication\/} {\bf #1}, #2}

\begin{opening}

\title{Recent Searches for Pulsars in Unidentified EGRET Error Boxes}
\author{M. Roberts}%$^{1,2}$}
\institute{%$^1$ 
Dept. of Physics, McGill Univ., Montr\'eal, QC, Canada\\
Center for Space Research, MIT, Cambridge, MA\\}
\date{} % DO NOT INSERT ANY DATE HERE !!!
\end{opening}

\begin{document}

%\oddpagefooter{\sf AGILE Pular Workshop}{}{\thepage}
%\evenpagefooter{\thepage}{}{\sf AGILE Pulsar Workshop}
\oddpagefooter{}{}{} % LEAVE AS IT IS !
\evenpagefooter{}{}{} % LEAVE AS IT IS !
\medskip  % LEAVE AS IT IS !

\begin{abstract} % LEAVE THIS COMMAND AS IT IS AND WRITE THE ABSTRACT IN THE FOLLOWING!

Pulsars discovered since the end of the $EGRET$ mission coincident 
with unidentified high-energy $\gamma$-ray sources are prime targets for
$AGILE$. Both general surveys and targeted observations have been successful
in finding energetic young pulsars in the plane of the Galaxy. 
For the latter, hard X-ray imaging followed
by deep radio searches of X-ray sources is a proven route. High-resolution
Chandra X-ray imaging has discovered several asymmetric pulsar wind nebulae 
associated with apparently variable $\gamma$-ray sources along the Galactic
plane. The unidentified sources at intermediate Galactic latitudes, with 
their much larger error boxes, make new X-ray studies impractical. A radio
search of a number of these sources using the Parkes multibeam system 
has resulted in the detection of several new binary pulsar systems. 
Whether these are potential $\gamma$-ray emitters has yet to be determined. 

\end{abstract}

\medskip

\section{Introduction}

The periods of all known $\gamma$-ray pulsars were originally discovered at 
lower energy wavelengths. Although it has been shown that pulsations could 
have been 
discovered from the Crab, Vela, and Geminga pulsars from just the 
$EGRET$ data without a priori knowledge of their pulse periods (Jones 1999),
all detections of $\gamma$-ray pulsations were achieved by folding the
data given an X-ray or radio ephemeris. 
While there is the hope that blind searches of $AGILE$ and $GLAST$ data may
result in some pulsar detections, the tendency of energetic young pulsars
to glitch and have timing noise and the planned default scanning mode of the
$GLAST$ mission will greatly complicate such searches.  Current radio 
or X-ray ephemerides and precise positions allow any amount of data to
be folded coherently without having to search through $P - \dot P$ space. 
They also make unnecessary the restriction of searches to observations of
a few months or less 
due to the large effective $\ddot{P}$ from timing noise and worries about
glitches. 
It is therefore highly desirable to detect potential 
$\gamma$-ray pulsars before the next generation of high energy $\gamma$-ray 
missions are launched. 

The $EGRET$ mission has left us a legacy of unidentified Galactic $\gamma$-ray 
sources at low and mid-Galactic latitudes, many of which will undoubtably 
turn out to be pulsars. 
In recent years, several new young pulsars have been found coincident with
unidentified $EGRET$ sources.
Some of these, such as the 69~ms PSR 
J1420$-$6048 in the Kookaburra (D'Amico et al. 2001), were discovered in 
the Parkes Multibeam Survey (Kramer et al. 2003). However, many young
radio pulsars are too faint to be detected in general surveys. Fortunately, 
young pulsars often reveal themselves as X-ray point sources or through 
the X-ray or radio emission of an associated wind nebula. In fact, 
about a third of the known pulsar wind nebulae (PWN) are associated with
known $\gamma$-ray pulsars or are coincident with unidentified $EGRET$
sources.

\section{Finding Potential $\gamma$-ray pulsars}

An unidentified $\gamma$-ray source can serve as 
a guide to lower energy searches, with the classic case being
the discovery of the Geminga pulsar through its X-ray emission 
(Halpern and Holt, 1992). In general, a strategy of hard ($> 2$ keV) 
X-ray imaging of a $\gamma$-ray error box followed by radio imaging 
and deep radio pulse searches has proven successful. 
This method has resulted in the discovery
of at least six pulsar wind nebula, three of which contain radio 
pulsars which were discovered subsequently 
(Halpern et al. 2001, Roberts et al.  1999, Roberts et al. 2002).     

The case of PSR J2021+3651 is an illustrative example. COS B first
detected a high-energy $\gamma$-ray source in this part of the
Cygnus region. Shallow radio pulse searches (Nice and Sayer 1997), soft X-ray
imaging and hard X-ray imaging using an early $EGRET$ position
failed to discover any sources of note (Mukherjee et al. 2000).
A 2-10 keV $ASCA$ image based on a revised position derived from 
$> 1$ GeV $EGRET$ data discovered the compact hard X-ray source
AX J2021+3651. The X-ray localization allowed a deep search for 
radio pulsations using the Arecibo telescope, resulting in the
discovery of the young, energetic (but radio faint $S_{20cm}\sim 0.1$~mJy) 
104~ms pulsar PSR J2021+3651 (Roberts et al. 2002). 
A follow up Chandra observation showed this to be embedded in a 
compact X-ray pulsar wind nebula with the standard torus and jet 
morphology (Hessels et al. 2003). 

Imaging of these type of nebulae may be important for constraining
models of high-energy emission. The orientation of the X-ray jets,
presumably aligned with the pulsar's spin axis, and that of
the torus, presumably aligned with the equator, tell us 
the viewing angle (Ng and Romani 2003). 
Combined with the polarization sweep (Radhakrishnan
and Cooke 1969) and possibly the pulse morphhology, the magnetic 
inclination angle can also be strongly constrained. Since models of high-energy 
pulsed emission generally depend on only the spin rate, magnetic field, 
and pulsar geometry, these observations should allow theorists to 
predict what $AGILE$ and $GLAST$ should see from these newly discovered
pulsars. 

\begin{figure}
%\epsfysize=6cm % fix the y-dimension and scales x-dim. to y-dim.
%\hspace{3.0cm} 
%\vspace{4.0cm}
%\epsfbox{vercellone_O_1fig1.ps}
\centerline{\epsfig{figure=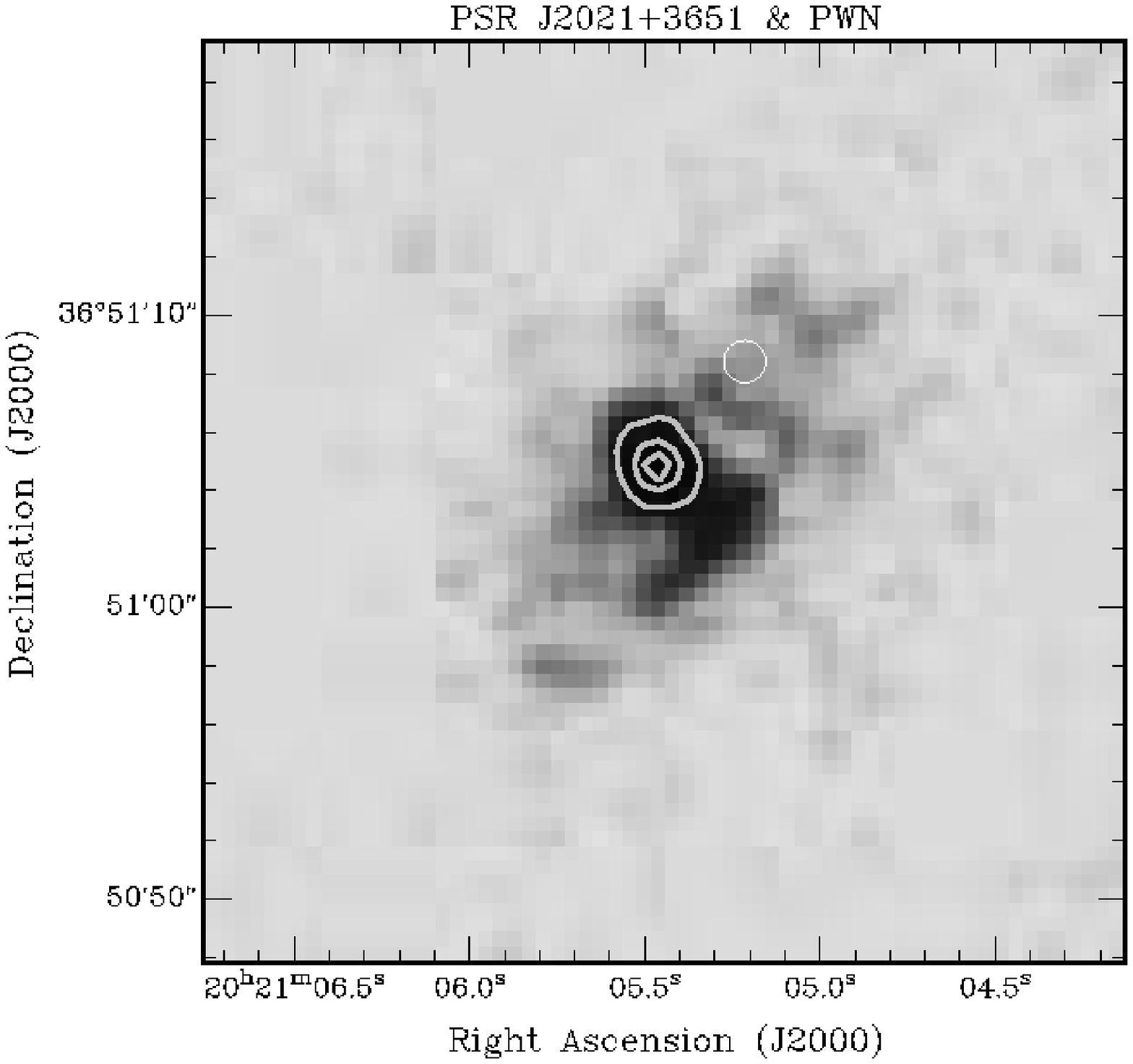,height=6cm} 
\epsfig{figure=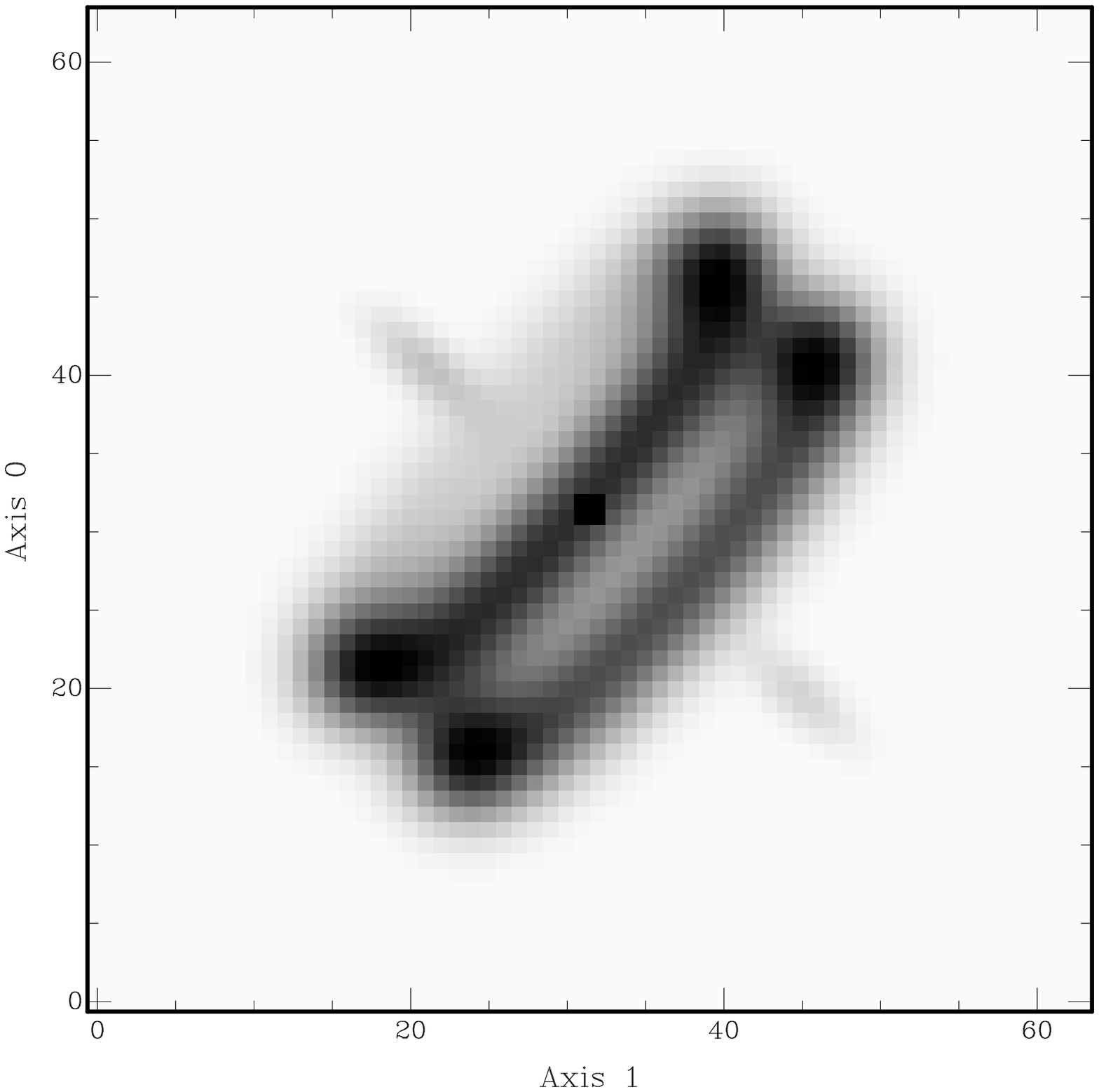,height=6cm} }
\caption[h]{Left: Chandra X-ray image of PSR J2021+3651. Right: Sample 
(not fit) model of torus plus jet for a specific geometry (courtesy C.-Y. Ng).}
\end{figure}

%
%Use the following minipage example for the inclusion of tables.

\vspace{.5cm} %TO ALLOW SUFFICIENT SPACE BETWEEN THE TEXT AND THE FIGURES

\noindent
\begin{minipage}{13.5cm}
\centerline{\bf Energetic Pulsars Recently Discovered in $EGRET$ Error Boxes}
\vspace{.5cm} %TO ALLOW SUFFICIENT SPACE BETWEEN THE TEXT AND THE FIGURES
\begin{center}
\begin{tabular}{|l|c|c|c|l|}
\hline

Pulsar &  log $\dot E$ & log $B$ &  D$^a$ & Ref.\\
      &erg/s&G&kpc&\\      

\hline

J2229+6114 &37.4 &12.3 &3.97 & Halpern et al. 2001 \\
J1420$-$6048 &37.0 &12.4 &5.63 & D'Amico et al. 2001 \\
J2021+3651 &36.5 &12.5 &12.4 & Roberts et al. 2002 \\
J1016$-$5857 &36.4 &12.5 &8.00 & Camilo et al 2001 \\
J1837$-$0604 & 36.3 & 12.3 & 6.4 & D'Amico et al. 2001\\

\hline
\end{tabular}
\end{center}
\smallskip
$^a$ Dispersion measure distances using the NE2001
model of Cordes and Lazio.

\end{minipage} 

\section{Pulsar Wind Nebula Associated with Variable $\gamma$-ray Sources}

The pulsed emission from the known $\gamma$-ray pulsars is very steady 
on timescales much longer than the pulse period. However, there is 
good evidence of at least one class of variable Galactic $\gamma$-ray sources
(McLaughlin et al. 1996, Nolan et al. 2003). $ASCA$ 2-10 keV images 
have been made of nearly all of the sources which are bright above 
1 GeV (Roberts, Romani and Kawai 2001). The brightest non-thermal 
source in the error boxes of the four Galactic GeV sources showing the
most evidence of variability appears extended even with the low resolution
of the $ASCA$ telescopes, suggesting they might all be pulsar wind
nebulae. Indeed, one of them is the known PWN around PSR B1853+01 in the
SNR W44 which has a trailing plume morphology in radio and X-rays 
(Frail et al. 1996, Petre et al. 2003). The other three sources have now 
been imaged in radio and with high resolution
X-ray telescopes. Two of them have radio nebulae with the spectral 
and polarization characteristics of PWN, both with 
bow-shock morphologies (Roberts et al. 1999, Braje et al. 2002). 
In X-rays, Chandra imaging of all four sources show a point source with a 
trailing jet type morphology.

In the table below, we list these four sources with their $V_{12}$ and
$\delta$ 
values from Nolan et al. 2003. $10^{-V_{12}}$ is the liklihood that 
$\delta < 0.12$, where $\delta \equiv \sigma/\mu$ (standard deviation over mean)
is a measure of how variable a source is with 0.12 being a conservative estimate
of the systematic variability of $EGRET$ data.  The typical time scale
of this variability is a few months, and the amplitude of the variability 
is on the order of the mean flux. Chandra observations of hard X-ray 
variability in the bright PWN 
around the Crab, Vela, and PSR J1811$-$1925 in SNR G11.2$-$0.3 have
established the dynamic nature of the emission from jet-like 
structures coming from pulsars (Hester et al. 2002, Pavlov et al. 2003, 
Roberts et al. 2003). Whether through beaming, magnetic 
field enhancements, or some other mechanism this emission can extend up to 
$\gamma$-ray energies is an open question.   

\vspace{.5cm} %TO ALLOW SUFFICIENT SPACE BETWEEN THE TEXT AND THE FIGURES

\noindent
\begin{minipage}{13.5cm}
\centerline{\bf PWN in Error Boxes of Variable $EGRET$ Sources} 
\vspace{.5cm} %TO ALLOW SUFFICIENT SPACE BETWEEN THE TEXT AND THE FIGURES
\begin{center}
\begin{tabular}{|l|c|c|l|}
\hline

PWN &  $\delta^a$ & $V_{12}^b$ & Ref.\\
\hline
Rabbit &  $1.03^{+0.80}_{-0.66}$ & 1.59 &  Roberts et al. 1999 \\
GeV J1809$-$2327 & $0.71^{+0.42}_{-0.25}$& 3.93 &  Braje et al. 2002 \\
GeV J1825$-$1310 & $0.88^{+0.57}_{-0.38}$ & 3.22 &  Roberts et al. 2001\\
PSR B1853+01 & $0.71^{+0.82}_{-0.43}$ &1.57 & Petre et al. 2003\\

\hline

\end{tabular}
\end{center}

\smallskip
$^a$ Variability magnitude parameter from Nolan et al. (2003) with 68\% confidence region
for $\delta \equiv \sigma/\mu $; $^b$ Variability determination statistic from
Nolan et al. (2003). $V_{12} = 1.3 $ rejects constant hypothesis at
95\% confidence level. \\

\end{minipage} 

\section{A Radio Survey for Pulsars in Mid-Latitude $EGRET$ Error Boxes}

While the exact nature of most of the low Galactic latitude sources remains
a mystery, in many cases plausible low-energy counterparts have been 
identified which can be confirmed by $AGILE$ and/or $GLAST$. At 
latitudes $|b| > 10^{\circ}$, only one or two non-Blazar candidate 
counterparts have been suggested for unidentified sources 
(eg. a neutron star for the singular high-latitude GeV source 3EG J1835+5921,
Mirabal et al. 2000).  At mid-Galactic latitudes, there is at least one
population of sources associated with the Galaxy which on average 
are weaker and have a significantly steeper spectrum than 
the unidentified sources along the Galactic plane (Hartman et al. 1999). 
Spatially, these may coincide with the Gould Belt, a local region
of recent star formation, and/or the Galactic Halo (Grenier 2002 and
elsewhere in these proceedings). Due to their proximity ($d \sim 50-300$~pc),
pulsars born in the Gould belt might appear $\gamma$-ray bright for longer 
periods of time and at larger off-axis angles than pulsars at typical
Galactic distances of a few kpc (Harding et al. 2003). It is also
possible that millisecond pulsars, whose Galactic scale height is much
greater than that of young pulsars, could form a halo population of 
$\gamma$-ray pulsars with a different typical spectra than the young pulsars
(Kuiper et al. 2000, Romani 2001). 

Practically speaking, these sources will be difficult to identify. The
typical 95\% confidence $EGRET$ error contour for these sources is 
$\sim 1.5^{\circ}$ across, much larger than the typically $< 1^{\circ}$ 
error boxes of the harder, low-latitude sources. The steep spectra 
of these sources mean that the great improvements in resolution projected
for $GLAST$, largely due to its good sensitivity to photons above 1
GeV which have relatively small PSFs, may not be nearly so great for these 
sources.  In fact, for the softest sources, $AGILE$ may do nearly as good a 
job at localization as $GLAST$. Since there are no wide-field
imaging X-ray telescopes operating in the 2-10 keV range,
the strategy outlined above is impractical with the current error boxes. 
A prime goal of $AGILE$ with respect to these sources is to localize them
to better than $30^{\prime}$ so as to be observable with a single X-ray
pointing by XMM. 

Radio pulse searches of the error boxes are practical and highly desirable. 
If there is a significant population of millisecond $\gamma$-ray pulsars, 
a priori knowledge of a current timing
ephemeris will be absolutely crucial to their detection as $\gamma$-ray
sources. The large search space in frequency and frequency derivative
required to detect millisecond pulsations in blind searches of $AGILE$ 
or $GLAST$ data will be computationally prohibitive and not very sensitive. 
In addition, the tendency of millisecond pulsars to be found in
binary systems further complicates searches. Even with slower, isolated
pulsars, a radio ephemeris would greatly facilitate the search for
$\gamma$-ray pulses. 

The Parkes Multibeam system makes deep radio pulse searches of large regions
of the sky at 20cm practical. The 13 feeds of the receiver are placed two beam widths 
apart on the sky, allowing a 4 pointing tesselation pattern to completely
cover an area of the sky $\sim 1.5^{\circ}$ across, matching well the 
typical mid-latitude $EGRET$ error box. We have completed a survey of 56 
unidentified $EGRET$ sources using the Multibeam receiver (Roberts et al. 2004). Each 
observation was $\sim 35$ min. long, with a sampling time of 0.125~ms. 
This is the same integration time as the highly successful Parkes Multibeam 
Galactic Survey (Kramer et al. 2003) and 8 times that of the Swinburne mid and
high-latitude surveys
(Edwards et al. 2001, Jacoby et al. 2003).  
In order to maintain sensitivity to pulsars in even tight binaries, 
we performed acceleration searches on all of the data using {\tt Presto}
(Ransom 2001).  The sources were selected using the following criteria:
$|b| > 5^{\circ}$ (so as not to overlap with the PMB plane survey), no 
probable blazar counterpart (Mattox, Hartman, Reimer 2001), declination
$< +20^{\circ}$ so as to be easily accessible to the Parkes telescope, and 
a 95\% confidence error contour which is well covered by the four pointing
tesselation pattern. 

Initial processing of the data has resulted in the discovery of
three new pulsars in binary systems and the redetection of a fourth. A
fifth binary pulsar in the surveyed area discovered in the Swinburne 
survey (Edwards and Bailes 2001) was not redetected. 5 binary pulsars
within the total  surveyed area is about twice what would 
have been expected from a simple extrapolation of numbers detected from
previous surveys. Due to the low number, whether this is statistically
significant is difficult to say. Since long term timing of these 
pulsars has only recently commenced, we do not yet know whether any of 
these new pulsars are energetically capable of producing the $\gamma$-ray
emission from their coincident $EGRET$ sources.   

\vspace{.5cm} %TO ALLOW SUFFICIENT SPACE BETWEEN THE TEXT AND THE FIGURES

\noindent
\begin{minipage}{13.5cm}
\centerline{\bf New Binary Pulsars in $EGRET$ Error Boxes} 
\vspace{.5cm} %TO ALLOW SUFFICIENT SPACE BETWEEN THE TEXT AND THE FIGURES
\begin{center}
\begin{tabular}{|l|c|c|c|l|}
\hline
Pulsar  & $P$ & $P_B$ &  D & Ref \\
      &s&d&kpc&\\
\hline
PSR J0407+1607 & 0.0257 & 669 &  1.32 & Lorimer et al. 2003 \\
PSR J1614$-$2238 & 0.00315 & 8.68 & 1.27 & Current Survey \\
PSR J1614$-$2315 & 0.0335 & 3.15 & 1.89 & Current Survey \\
PSR J1744$-$3924 & 0.1724& 0.19 & 3.05 & Current Survey + PMB\\
PSR J1745$-$0952 &0.0194 & 4.94 & 1.83 &Edwards and Bailes 2001 (not detected)\\
\hline

\end{tabular}
\end{center}

\end{minipage}

\vspace{.5cm} %TO ALLOW SUFFICIENT SPACE BETWEEN THE TEXT AND THE FIGURES

We have also discovered three new isolated pulsars. In addition, we
redetected 6 out of 7 previously known pulsars. Again, we do not yet have 
measured spin-downs for the new pulsars, and none of the
previously known pulsars with published period derivatives
appear energetically capable of producing 
$\gamma$-rays. We are not certain why we have detected so few isolated,
slow pulsars. However, at the sensitivity threshold of this survey, 
the low-frequency RFI in our data is quite problematic. For slow, 
low DM pulsars expected from the Gould belt, the pulse dispersion at 20cm 
is not great enough to distinguish pulsar signals from man-made pulsed signals.
While we may be able to extract a few more slow pulsars from this data, 
it will probably require a longer wavelength survey to have good 
sensitivity to pulsars in the Gould belt.

\vspace{.5cm} %TO ALLOW SUFFICIENT SPACE BETWEEN THE TEXT AND THE FIGURES

\noindent
\begin{minipage}{13.5cm}
\centerline{\bf Isolated Pulsars in $EGRET$ Survey Error Boxes}
\vspace{.5cm} %TO ALLOW SUFFICIENT SPACE BETWEEN THE TEXT AND THE FIGURES
\begin{center}
\begin{tabular}{|l|c|c|c|l|}
\hline
Pulsar  & $P$ & $log \dot E$ &  D & Ref \\
& s & erg/s & kpc & \\
\hline
J1632-10 &  0.7176  & - & $ > 50$ & Current Survey \\
J1636-1509 & 1.1794 & - & 1.99 & ATNF not detected \\
J1650-1654 & 1.7496 & 31.4 & 1.47 & ATNF \\
J1725-07 & 0.2399 & - & 1.69 & Current Survey \\
J1741-2019 & 3.9045& 31.0 &  1.72 & ATNF \\
J1741-3927 & 0.5122& 32.7 & 3.21 & ATNF \\
J1800-01 & 0.7831& - & 1.62 & Current Survey \\
J1821+17  &  1.3667  & - & 3.26 & ATNF \\
J1832-28  & 0.1992& - & 3.32 & ATNF \\
J1904-1224 & 0.7508& 31.8 & 3.37 & ATNF \\
\hline
\end{tabular}

\end{center}

\end{minipage}

\section{Conclusion}

Both large scale radio surveys and deep radio searches targeting
X-ray sources have resulted in the discovery of interesting 
new pulsars coincident with unidentified $EGRET$ sources. 
X-ray and radio imaging of $EGRET$ error boxes has also
revealed the presence of energetic new pulsars through 
the emission of associated wind nebulae. Radio pulse searches 
of other X-ray sources and ultra-deep observations of the 
currently known PWN are planned or underway. We have also 
begun a northern extension of our mid-latitude survey using 
the GMRT. The goal is to discover and start timing as many potential 
$\gamma$-ray pulsars as possible before the launch of $AGILE$.    
We also hope to obtain deep, high resolution images of PWN 
to determine accurate morphologies which can constrain pulsar
geometry allowing case by case predictions of what will be
seen with the next generation of $\gamma$-ray satellites. 

Besides detecting pulsations, goals for $AGILE$ include 
confirming $\gamma$-ray variability and  possibly correlating it
with X-ray variations. In addition, every effort should be made
to constrain the mid-latitude source error boxes to be within the
field of view of imaging X-ray satellites.  This will allow counterpart
searches to commence before or concurrent with the launch of $GLAST$. 

\acknowledgements

I thank C.-Y. Ng for the sample PWN model image. I also thank Jason Hessels,
Cindy Tam, Vicky Kaspi, and Scott Ransom for comments.

\end{document}